# Electrochemically-driven formation of Intermetallic $Cu_3ZnLi_2$ alters Li-transport in nanostructured bimetallic battery anode

Eric V. Woods[1,*], Xinren Chen[1], Yuwei Zhang[1], J. Manoj Prabhakar[1], Patricia Jovičević-Klug[1], Matic Jovičević-Klug[1], Mahander P. Singh[1], Yujun Zhao[1], Siyuan Zhang[1], Stefan Zaefferer[1], Jian Liu[2], Yug Joshi[1,*], and Baptiste Gault[1,3,*]

1. Max Planck Institute for Sustainable Materials (MPI-SusMat, formerly Max-Planck-Institut für Eisenforschung GmbH (MPIE)), Max Planck Straße 1, 40237 Düsseldorf, Germany
2. School of Engineering, Faculty of Applied Science, The University of British Columbia, Kelowna V1V 1V7, Canada
3. Department of Materials, Royal School of Mines, Imperial College London, SW7 2AZ, London, UK

*Corresponding authors: e.woods@mpi-susmat.de, y.joshi@mpi-susmat.de, b.gault@mpi-susmat.de

Funding: Supported by the German Research Foundation (DFG) Leibniz Prize and other agencies (see Acknowledgements).

Keywords: lithium-metal batteries, brass current collectors, Laves phase, nanocrystalline, dendrite suppression, lithium ion batteries

## Abstract

The role of Li-based batteries in the electrification of society cannot be understated, however their operational lifetime is often limited by the formation of dendrites, i.e. the localised deposition of Li that can cause shorts between the two electrodes leading to the failure of the battery. Nanocrystalline bimetallic current collectors can be used for anode-free Li-metal batteries, with improved Li plating and limited or suppressed formation of dendrites. Here, we demonstrate that the microstructure of an α-Brass current collector, Cu 63% Zn 37%, used in an anode-free Li-metal battery evolves during cycling. It initially had a nanocrystalline deformation layer approximately 80 nm in thickness after polishing. After 100 cycles, the initial deformed brass layer was partially converted to a ternary Laves phase $Cu_3ZnLi_2$ within a nanocrystalline brass matrix that grew to 200 – 250 nm in thickness. Upon Li stripping, the phase partially decomposes electrochemically, but what remains can sequester Li, thus forming "dead Li" thereby contributing to capacity loss. We propose a mechanism for the microstructural evolution including dynamic recrystallization and phase formation. Since this ternary Laves phase emerges during electrochemical cycling alone, binary alloy current collectors must be assessed for metastable ternary phase formation under different cycling conditions to either stabilize and exploit such phases or electrochemically fully strip them.

# Introduction

The formation of Li dendrites is a primary cause of catastrophic failure in lithium-ion batteries (LIBs), particularly Li-metal batteries (LMBs) [1–5]. Different approaches for suppressing Li dendrite formation have been explored, e.g. by improving Li plating uniformity onto the current collectors of LMBs, including electrolyte chemistry [6–9], improving the solid electrolyte interface layer (SEI) stability [10], changing pressure [11], coatings [9,12,13], and alloyed current collectors [14,15]. Structured three-dimensional collectors have good stability, as reviewed in [16,17]; in that category, porous current collectors have been developed, such as Cu foam [15] and porous Cu produced via dealloying methods [18,19]. Nano-grained Cu current collectors formed by advancing rolling techniques have been shown to increase nucleation sites and improve plating efficiency [20]. Cu alloys, such as CuZn, have been used as current collectors [21] as well as porous CuZn or CuZn-based materials [22–24] and CuSn [25].

Alloy anodes of various compositions have been tested for Li-metal batteries, some of which alloy with Li during plating, such as In, Ag, and others [18,26–28] or segregate as bilayers [11]. Typically, such systems evolve novel metastable phases during that alloying process [29,30,30]. Two-component Li alloys have been explored in anode-free Li metal batteries to improve plating, such as LiZn [23,31,32] and LiSn [32–34]. Normally, Cu and Li are immiscible at room temperature according to the phase diagram, but at the nanoscale, even normally immiscible elements can potentially alloy or form metastable phases. As an example, electrochemically lithiated CuO nanowires can form amorphous $CuLi_x$ alloys [35].

Ternary Cu-Zn-Li alloys have been explored as anode material for LMBs, such as $Li_{60}CuZn_5$ formed via melting Cu, Zn, and Li metals together under inert atmosphere [36]. This latter material formed CuZn "lithium-inactive" and $CuZn_5$ "lithium-active" phases after Li stripping [36]. The bulk ternary alloy $Cu_3ZnLi_2$ was formed by smelting, e.g. adding Cu and Zn powder to molten Li under inert atmosphere and used as an anode in LMBs [37]. Separately, the $Cu_3ZnLi_2$ phase was formed by pouring molten LiZn over Cu foil under inert atmosphere to form that phase for battery applications [37–39]. Likewise, nanowires and thicker layers of this alloy and similar Li-Cu-Zn phases can be formed by pouring molten CuLi onto Zn or by heating a Li film plated onto an ultrathin Zn layer on Cu [40,41]. However, formation of a ternary phase due to the alloying process during cycling has not yet been observed.

Precise characterization of Li-containing phases is essential for understanding Li distribution and transport within the current collector, along with their evolution. They can indeed form irreversible, trapped "dead Li" that no longer contributes to the cycling and hence lowers the battery capacity. Further, changes in the fraction of Li transported into and sequestered within that layer leads to local volumetric changes with associated stress fields, which could lead to a variety of defect mechanisms, affecting the structural and microstructural integrity, leading to cracking and detachment of fragments of the material and hence to capacity loss.

Here, we report that an initially 80 nm thickness nanocrystalline surface layer on a polished CuZn37 sample (Cu: 63%, Zn: 37%) used as the current collector in a Li-metal half-cell. After 100 cycles, the half-cell was opened to check Li plating uniformity and analyze the microstructural changes. During cycling, the nanocrystalline metallic interlayer partially converts to the ternary Laves phase, $Cu_3ZnLi_2$. The $Cu_3ZnLi_2$ phase remains present in the stripped electrode. The layer grew to approximately 200 – 250 nm thickness and exhibited significant dynamic recrystallization; the Li-stripped collector still retained about 10 – 15 at.% Li, contributing to capacity loss. Below an abrupt transition where Li concentration went to zero, a 500 nm-deep heterogeneously Zn-depleted zone undergoing dynamic crystallization emerged, indicating extensive and dynamic microstructural evolution. Formation of metastable ternary phases in Li-alloying anodes or current collectors can affect capacity retention even over a relatively low number of cycles (e.g. 100 here). We demonstrate that electrochemical mechanisms alone can drive metastable phase formation, e.g. non-stoichiometric Cu-Li-Zn compositions beyond the thermodynamically room-temperature stable Laves phase $Cu_3ZnLi_2$ comparable to other bimetallic systems [39]. Additionally, that alloying causes dynamic interfacial migration hundreds of nanometers below the initial current collector surface. These results showcase the importance of understanding microstructural evolution of battery electrodes to better understand their performance, and to guide selection of current collectors for "anode free" and alloy anodes and their processing.

## Results

### Capacity retention and GIXRD of plated versus stripped brass current collectors

Coin cells were assembled in a half-cell configuration (Li|brass) with the brass current collectors having an initial 80 nm thick nanocrystalline layer, as shown in **Figure S1**, and cycled 100 times. **Figure 1(a)** illustrates the cycling performance of the Li|brass cells over the first three cycles, with the first cycle charging showing a dip below zero potential in the lower left corner, which indicates that the overpotential for the plating of lithium on the pristine surface. This could also indicate that some irreversible alloying occurs between Li and Zn. The charge capacity for increasing cycle number is plotted in **Figure 1(b)** for the nanocrystalline brass current collector. It shows a decline up to around cycle 40, then somewhat stabilizes. The nanocrystalline brass current collector's Initial Coulombic efficiency (ICE) was 81% versus 65.3% for Cu foil in a control cell, as indicated **Figure 1(c)**, indicating Li sequestration already during the first cycle, albeit mechanisms differ. Li plated on Cu foil becomes inactive through SEI formation and side reactions, whereas Li alloys with brass as detailed in the following [42–44].

After 100 cycles, coin cells were disassembled to perform grazing incidence X-ray diffraction (GIXRD) on both the Li-plated, electrochemically stripped brass and Cu current collectors. The results are shown in **Figure 1(d)**, where the peak at 22° is the strongest and it pertains to the $Cu_3ZnLi_2$ Laves phase; in the stripped sample, the lower intensity of the peak indicates that the volume fraction of $Cu_3ZnLi_2$ was reduced, and the ratio of the primary Cu peak at 42° changes from approximately 6:1 in plated phase to about 10:1 in the stripped phase. Rietveld refinement of the GIXRD data shows that the observed $Cu_3ZnLi_2$ volume fraction decreased from 8.1% when plated to 6.5% when stripped, while Cu rises from 45.2% when plated to 60.3% when stripped, as summarized in **Supplementary Table T1.** Density functional theory (DFT) based calculations suggest that the Laves phase is thermodynamically stable at room temperature [38,39], explaining the only slight change in volume fraction of the Laves phase after Li stripping. This also suggests that its formation is only partially electrochemically reversible. Additional component phases, Rietveld R values, goodness-of-fit given as $\chi^2$ are provided in the **Supplementary Information** in **Table T1.** Note that while Li remained on the plated collector, the X-ray intensity attenuation is likely minimized (e.g., < 25%), because Li has the lowest X-ray scattering cross-section [45]. Additional peaks corresponding to LiOH & $Li(OH)_2$ formation from Li reaction with atmospheric moisture are present, and the Cu signal is difficult to deconvolve from the remaining low-Zn brass [46].

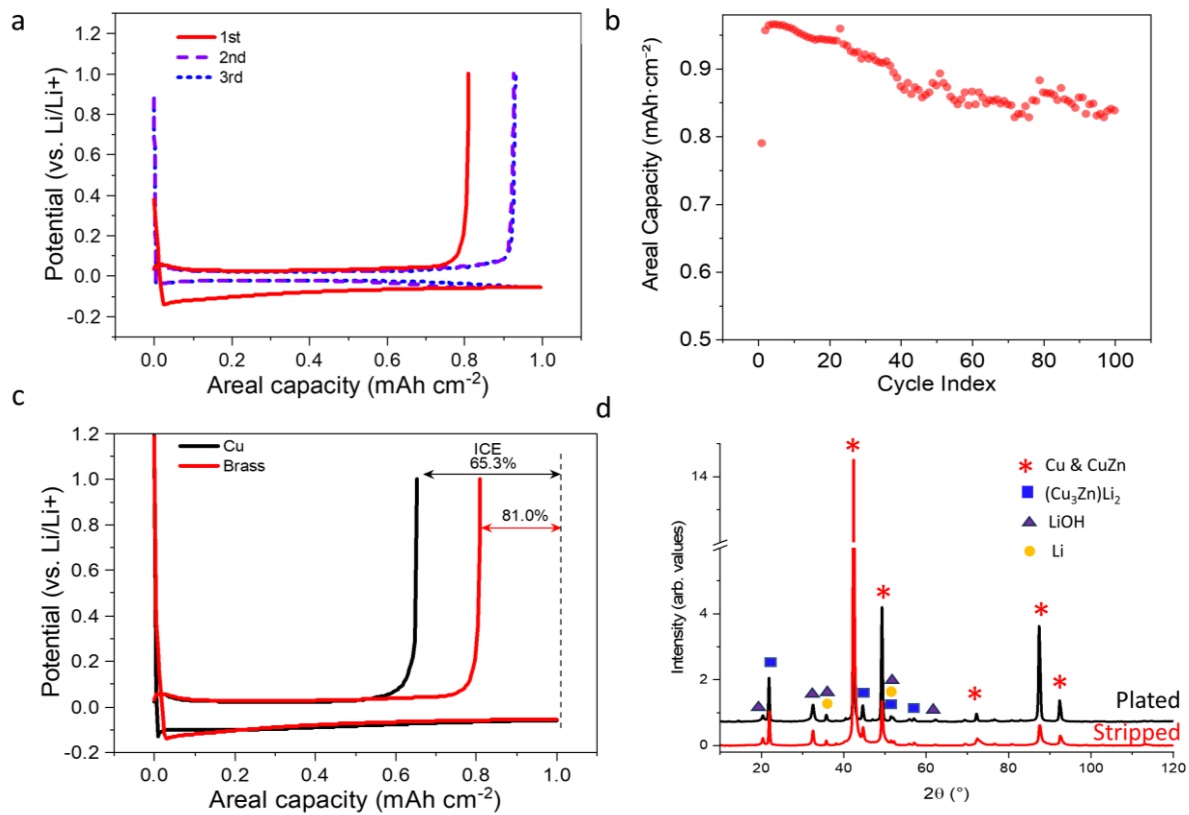

**Figure 1.** (a) First 3 cycles of Li|brass, with the slight indication of Li-Zn alloying, based on the dip of the green line below 0 V highlighted by the black box, showing overpotential around -0.1 V (b) Areal capacity change of Li|brass over 100 cycles (c) comparison of first-cycle capacity of Li|brass versus Li|Cu foil control (d) Stacked XRD patterns of plated 100X cycled Li-plated versus stripped 100X-cycled brass current collector

## Li-plated brass current collector after cycling

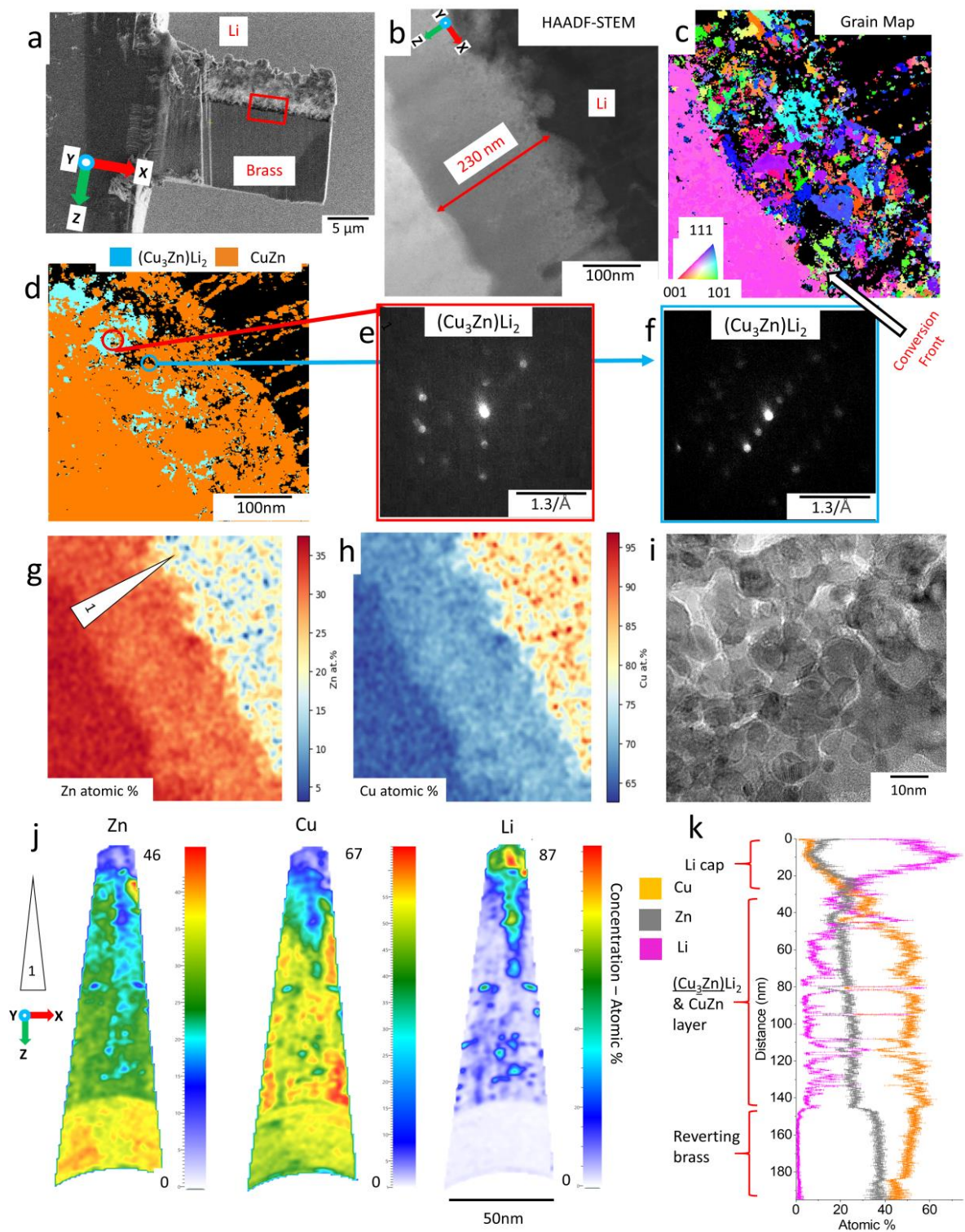

**Figure 2.** Plated 100X cycled brass current collector (a) FIB-prepared lamella (b) HAADF-STEM interlayer view (c) X-direction (scan direction along X-axis) grain map (d) indexed phase distribution 4D-STEM (e) diffraction pattern Laves phase (f) diffraction pattern second site indexed as Laves phase – comparatively indexed in **SI 4** (g) STEM-EDX Zn distribution (at%) with functionally equivalent APT region noted as 1 (h) STEM-EDX Cu distribution (at.%) (i) TEM overview grain view (j) Zn, Cu, and Li distribution APT 2D XZ center slice directed down from plated Li surface (k) overall XZ projection Z-composition profile with labelled phases

After cycling, scanning electron microscopy (SEM) imaging of a lamella extracted at cryogenic temperatures by using focused-ion beam (FIB) (see Methods), **Figure 2(a),** reveals an approximately 4 µm thick uniform Li layer on the plated current collector. Cryogenic temperatures during preparation and imaging are necessary to avoid affecting the Li-distribution. The nanocrystalline interlayer was observed via high angle angular dark field (HAADF)-scanning transmission electron microscopy (STEM) image in **Figure 2(b)**, with the lighter colour illustrating an average lower atomic number, e.g. Li-intercalated regions. 4D-STEM was used to provide a grain orientation map, defined along the horizontal X-scan direction with a -32° rotation, plotted in **Figure 2(c)**. The corresponding phase map in **Figure 2(d)** confirms that the nanocrystalline layer contains CuZn and an estimated 20% volume fraction of $Cu_3ZnLi_2$. As visible in **Figure 2(d)**, the small size, e.g. 10 nm scale, of the deformed and overlapping grains made unambiguous indexing of overlapping diffraction patterns difficult. Two examples of indexed regions are shown in **Figure 2(e)** and **Figure 2(f);** examples of the closest diffraction patterns in indexing are shown in **Figure S2**. STEM-Energy dispersive X-ray spectroscopy (EDX) maps for Cu and Zn are plotted in **Figure 2(g)** and **Figure 2(h),** with respective an average interlayer composition of 75 at.% and 25 at.%. A higher resolution TEM image in **Figure 2(i)** illustrates that the grain size is on the order of 10 nm; enlarged view in **Figure S3**. STEM - electron energy loss spectroscopy (EELS) spectrum imaging was performed at room temperature; one Li EELS scan, shown in **Figure S4**, shows that the Li is in the LiOH state. Therefore, vacuum transfer with air protection is required to preserve Li; since APT was performed with that transfer, it is the better solution for imaging Li.

This motivated the use of atom probe tomography (APT) with full cryogenic specimen transport (cryo-APT) to map the Li distribution [47,48]. A first region-of-interest (ROI), functionally comparable to the one marked by the triangle # 1 in **Figure 2(g),** was analyzed and **Figure 2(j)** displays the XZ-2D compositional maps for Zn, Cu, and Li evidence intermixing of elements in the plated layer in an 8 nm thick layer, along with Li-rich pockets in the nanocrystalline region. These are above what we term a "conversion front", where the Li content drops to zero, and with a compositional gradient evolution suggestive of Zn diffusion, or a leaching process, which can explain the modification of the grain structure in the material bulk below this depth, as will be explained further below. To be clear, there is a Zn gradient, with respect to nominal composition vs Cu in the original material. This is also clearly evident below via STEM-EDX where the relative composition of Cu and Zn are changed in the conversion zone. Two sample ROIs across Li-rich regions in the middle region are shown in **Figure S5(a-b);** the 3D views and 1D profiles illustrate the presence of Li-rich metastable phase **Figure S5(a)** and Cu-Li equiatomic composition in **Figure S5(b).**

**Figure 2(k)** plots a 1D Z-axis composition profile. The Cu:Zn ratio is approximately 3:1 as expected for the Laves phase throughout the nanocrystalline region. There are large fluctuations in the Li composition, with what appear to be very high concentration Li pockets. Because Li field evaporates faster than the surrounding matrix, trajectory aberrations could cause intermixing and the composition is likely even higher [49]. These pockets appear are located in regions with lower Cu composition; these may be associated to grain boundaries (GBs) or triple points in the nanocrystalline layer. Below this layer, the Li composition is below the level of background, and there are gradients for Cu and Zn progressively reverting to the composition of the original CuZn37 alloy.

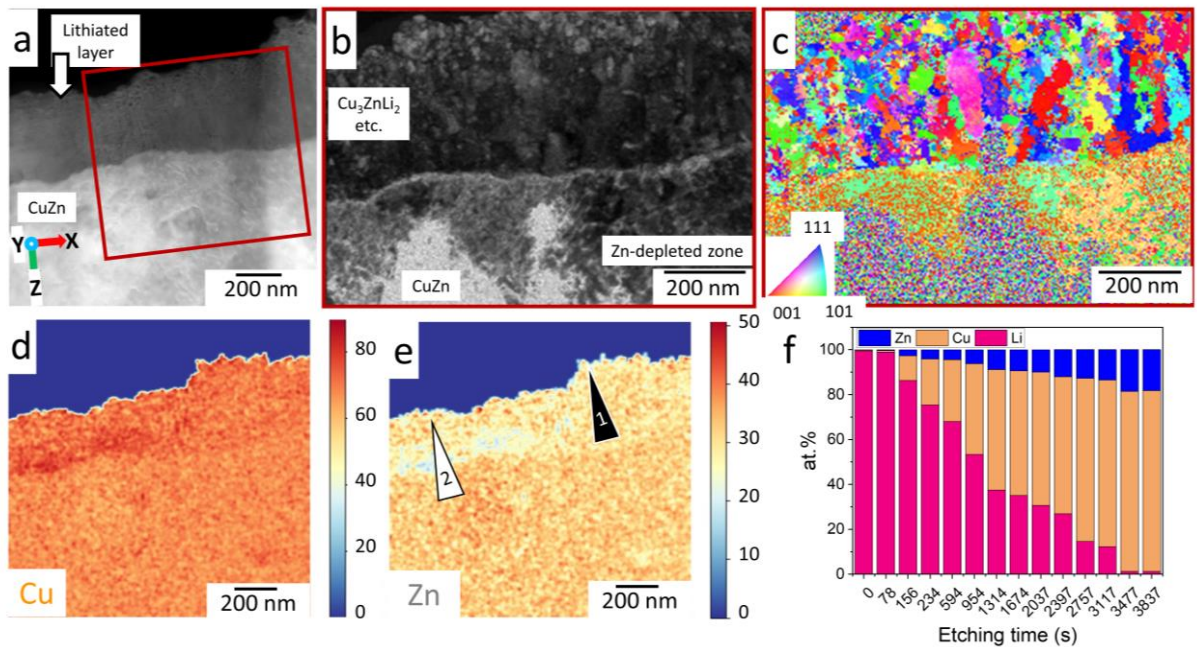

**Figure 3.** Stripped 100X cycled brass current collector. (a) HAADF view of the lamella, showing the lower atomic number interlayer on the top (e.g. lithiated brass) (b) enlarged DF-STEM image of red box in (a) analyzed in (c) 4D-STEM grain size map, which indexed as CuZn only, X-direction (arbitrary scan direction with beam) (d) STEM-EDX Cu distribution (at.%) (e) STEM-EDX Zn distribution (at.%) with functionally equivalent APT specimen positions marked, e.g. homogenous versus heterogeneous tips (f) XPS depth profile Zn:Cu:Li composition, sputtered to Li removal.

Li-stripped brass current collector after cycling

A sample from the electrochemically stripped 100X cycled brass electrode was similarly prepared and examined using the same experimental protocol. Peaks in the GIXRD measurements in **Figure 1(d)** confirm that the $Cu_3ZnLi_2$ phase was stable after delithiation after 100 cycles, although the volume fraction was reduced as in **Supplementary Table T1**. **Figure 3(a)** provides an overview STEM HAADF image. The dark field TEM image from a selected area marked by a red rectangle is displayed in **Figure 3(b)** and reveals a grain size comparable to, if not smaller than, the plated, cycled brass current collector. The 4D-STEM image in **Figure 3(c)** further corroborates this grain size estimation, with the orientation scan direction in the X-direction with no rotation. STEM-EDX in **Figure 3(d–e)** illustrates significant heterogeneity in the Cu and Zn concentrations within the interlayer.

Composition profiles from the X-ray Photoelectron Spectroscopy (XPS) ion-sputtered depth profile are plotted in **Figure 3(f)** along with further interpretation of the chemical signatures in **Figure S6**. Briefly, at the micro-scale, there is a definite Li vertical compositional gradient and slight Zn gradient, which was seen in the APT dataset for the plated collector. The Li concentration decreases to below the XPS detection limit (estimated between 1 – 10% for Cu/Zn matrices) before depth profile termination, indicating profiling through the entire 250 nm intermetallic layer [50]. Overall, this confirms that Li concentration declines with depth, consistent with previous findings in multiple Li-alloy anode systems, including Li-Zn [30]. This effect partially arises from differential elemental etch rates; while the local, nanoscale material heterogeneity and ternary phase probably minimizes this effect, residual electrolyte could cause differential etching and induce topographic effects [51,52]. Nonetheless, a compositional gradient is evidenced qualitatively. **Figure S6** also shows Li-halide bonding that decreases in intensity at increasing depth. Li-F formation results from reactions with the LiPF$_6$ salt within the liquid electrolyte.

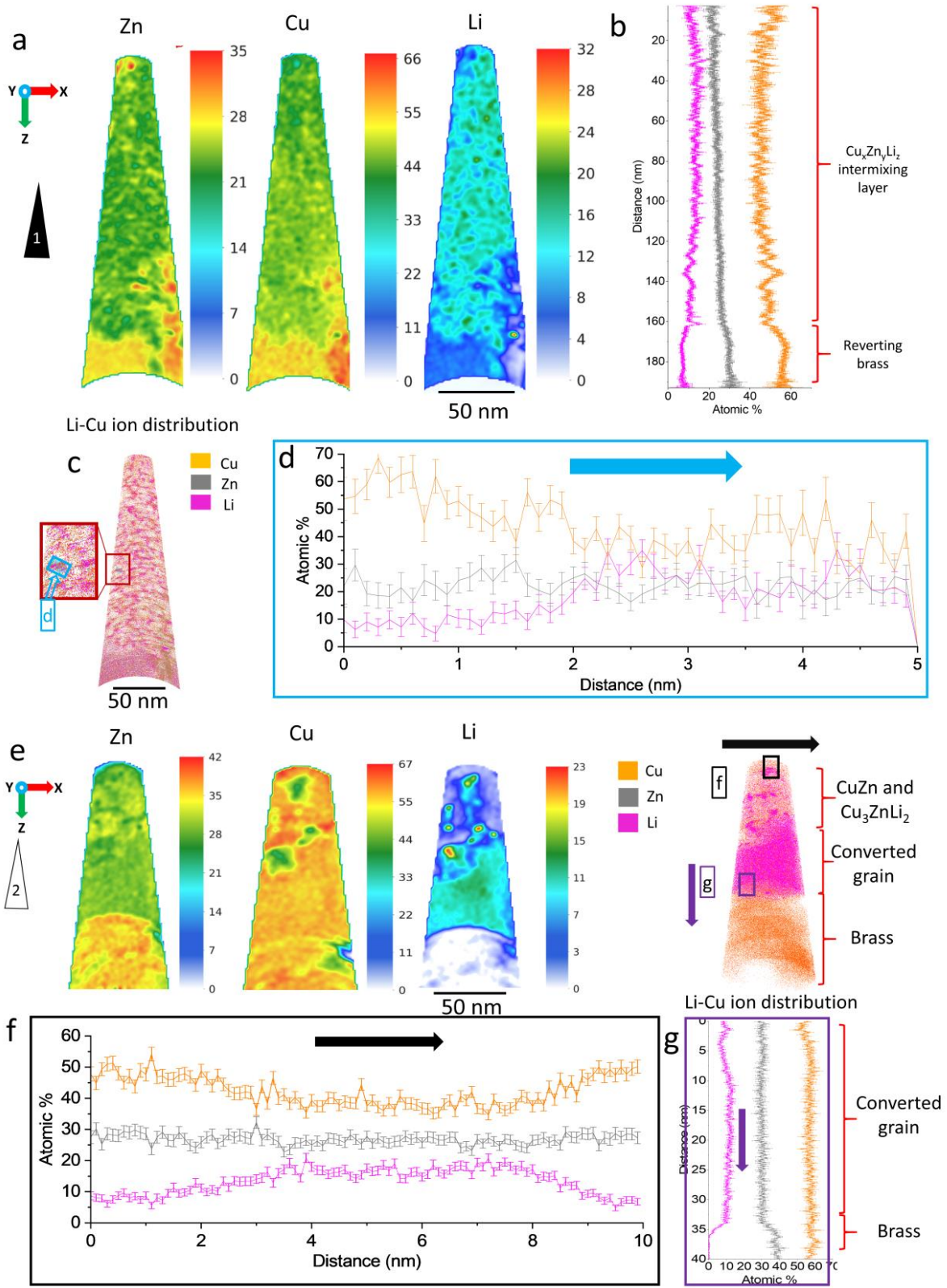

**Figure 4**: (a) APT 2D XZ center slice Zn, Cu, and Li composition maps (b) overall 1D center-slice composition profile for Zn, Cu, and Li composition, showing overall homogeneity, with labelled regions (c) Li-Cu ion visualization in center homogenous region XZ slice with a selected small region magnified in the red box, with a further blue box showing location of Li-enriched region noted in (d) overall 1D

projection Z-composition profile for blue box region with the left side Cu-rich grain and right side Li-rich nanoscale region (e) Zn, Cu, and Li APT 2D XZ center slice heterogeneous specimen and Li-Cu ion distribution with labelled overall regions and two specific ROIs shown below (f) 1D elemental profile in X direction for black box in the Li-Cu ion distribution map in (e), scan direction indicated with black arrow (g) 1D Z-composition profile for purple box showing the Li rise and abrupt fall in the converted grain as it reverts to brass, purple scan arrow denoting direction

A cryo-APT specimen from a compositionally homogenous region, equivalent to that denoted as 1 in **Figure 3(e)** was analysed. **Figure 4(a)** provides compositional maps for Zn, Cu, and Li within a 2D 15-nm-thick XZ slice through the APT reconstruction. The 1D Z-direction composition profile in **Figure 4(b)** shows an average Cu:Zn:Li ratio no longer consistent with the $Cu_3ZnLi_2$ intermetallic, but **Figure 4(c)** displays the section of the point cloud from the same slice that further evidences that Li appears distributed primarily in between Cu-rich regions. Yet a 1D compositional profile across a Li-rich particle, plotted in **Figure 4(d)**, illustrates a local increase in Li and decrease in Cu, approaching the stoichiometric Laves phase composition. We argue that these observations explain the Cu-rich "skeletons" previously reported [36]. Below, the Li composition tapers off slightly before the end of the dataset, indicating that the reaction front was just reached. This dataset evidences complex elemental partitioning behavior and the formation of nanoscale Laves phase particles.

A second cryo-APT specimen from a region compositionally equivalent to the one denoted as 2 in **Figure 3(e)** was analyzed. **Figure 4(e)** maps the Zn, Cu, and Li distributions in 2D elemental XZ-maps within a 12-nm-thick slice across the reconstructed data. A similar inverse relationship between the concentrations of Cu and Li is again visible in the top 50–60 nm. That top region, denoted as CuZn and $Cu_3ZnLi_2$ in **Figure 4(f)**, has a similar structure compared to the region under the Li cap in the Li-plated specimen in **Figure 2(k)**, with two comparable ROIs across Li-rich regions in **Figure S5(a-b).** Below that follows a region in which the composition appears homogenous and with an average composition consistent with the 1D Z-compositional profile in **Figure 4(g)**. Given that large grains exist in the intermetallic, this region is interpreted to be a fully converted grain with a metastable homogenous composition like $Cu_3ZnLi_2$. These phases that did not decompose fully during stripping and the remaining Li within these regions corresponds to "dead Li" from the perspective of the battery, contributing to the loss of capacity.

Finally, the Li composition abruptly drops to zero at a "conversion front" where the nanocrystalline region ends and the analysis reaches the brass substrate. However, Zn depletion and nanocrystalline formation clearly continues below the "conversion front", and there is substantial heterogeneity for the Zn composition consistent with Zn-depletion forming nanoscale regions under the conversion front. In combination with the 4D-STEM, **Figure 3(c)**, that indicates multiple grains extending into the bulk, we can deduce that Zn depletion can be accompanied by the formation of new grains and possibly dynamic recrystallization even below the conversion front in the area above the reaction front.

## Discussion

Our results demonstrate that during Li electrochemical plating / stripping cycles on a defect-rich, nanocrystalline brass surface, the ternary Laves phase $Cu_3ZnLi_2$ forms and partially decomposes. This explains why the initial Li-Zn alloying in the first cycle is seen as only partially reversible [31], and the capacity change over the first 100 cycles, particularly in the last 50 cycles. The ternary Laves $Cu_3ZnLi_2$ phase was reported to promote superior Li plating and suppress dendrites [37–39], which is further supported by our experiments. This could be related to faster incorporation of Li within the Laves phase compared to the brass itself, thereby facilitating Li transport and plating homogeneity. Prior work shows that nanocrystalline materials or coatings improve Li nucleation and plating as well as dendrite suppression [31,53–56]. A schematic representation of the mechanism of microstructural evolution over successive plating/stripping cycles is schematically depicted in **Figure 5**.

**Figure 5(a)** depicts that during the first plating cycles, the Li plates on the brass and diffuses along grain boundaries, starting to form small pockets of Li under the surface. Li diffuses three orders of magnitude faster along grain boundaries than bulk in Cu [57]. The plated Li surface lies above the Cu-Zn interlayer enriched in $Cu_3ZnLi_2$, and the brass substrate lies underneath. As Li plates, Zn atoms in the underlying brass rapidly diffuse upwards, contributing to the formation of the $Cu_3ZnLi_2$ ternary phase. Note that the lithiation of other alloy system, e.g. $Cu_6Sn_5$ or Zn-Sn alloy also leads to the formation of ternary intermetallic $Li_{13}Cu_6Sn_5$ [58,59] and $Li_3Zn_2Sn_4$ [14]. So, our observations may also apply more generally. Here the kinetic limiting factor is likely to be the Zn diffusion. First, Zn was reported to diffuse typically three times faster than Cu in brass [60–64]. Second, there is a dense grain boundary network in the nanocrystalline layer, as confirmed by TEM in **Figure 2(i),** and Zn diffusion along grain boundaries in pure Cu at 700 °C was approximately $10^6$ higher than through bulk [65]. This unbalanced atomic flux creates a downward counter-flow of vacancies, which is known as the Kirkendall effect. The Zn depletion leads to tensile stresses, associated to the lattice volume change, in two ways: between the dezincified layer and the substrate, and intergranular stress from frustrated volume change. The vacancies can assist further diffusion, but also accumulate and progressively form voids that relieve stress and provide preferential regions for Li to nucleate and migrate, as in these pockets observed by APT specimen, **Figure 2(j)**. Finally, Zn is likely the primary mobile metal species within the material, and it therefore controls early-stage kinetics of the Li-Cu-Zn interlayer growth, thereby controlling reaction- and conversion-front progression.

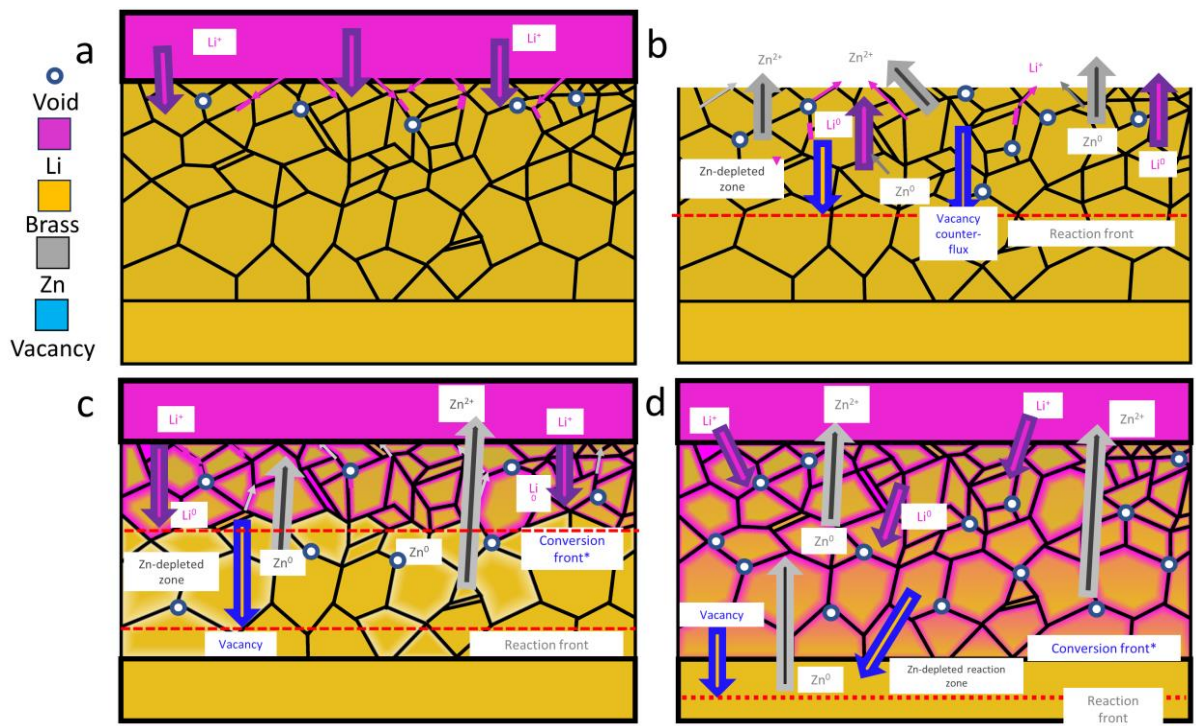

**Figure 5.** Model of atomic and ionic fluxes during plating and stripping for interlayer (diagrams not to scale; *interlayer is approx. 250 nm, Zn-depleted reaction zone is 500+ nm*). (a) initial Li plating onto nanocrystalline brass (b) Li stripping after first cycles highlighting the conversion front (c) Li plating after 40 cycles showing conversion and reaction fronts and undergoing dynamic recrystallization, with Li enrichment and incorporation into grain noted in pink gradient and Zn depletion noted in yellow gradient around grains below the conversion front (d) plating after 100 cycles illustrating the fully converted nanocrystalline layer and continuing Zn depletion causing further nanocrystalline layer formation and showing dynamic recrystallization. Note: grain coarsening not shown for simplicity.

In **Figure 5(b)**, after several stripping cycles, when the Li is stripped, the Zn also diffuses out into the electrolyte; the Zn has significantly depleted from the near-surface region. The Zn-depleted layer creates a strong tensile stress gradient with the underlying brass layer [66–68]. This depletion is illustrated in the compositional change at the bottom of **Figure 2(j)** in the "reverting brass" region as well APT in **Figure 4**, as well as indirectly in **Figure 3(b-c)**, with the area underneath the nanocrystalline layer converting to crystalline and heterogenous Zn concentration under that layer in Zn STEM-EDX in **Figure 3(e).** The Zn depletion creates a counterflow of vacancies that migrate deeper into the nanocrystalline brass region [69], along with a stress gradient from the Zn-depleted grains being unable to change their lattice constants [64,70,71], because they are constrained by their neighbours, as stated in Vegard's law [72]. As seen in the APT data in both **Figure 2(j)** and **Figure 4(e)**, there is a conversion front where the nanocrystalline region stops and the Li composition drops to zero. In **Figure 5(c)**, after some dozens of cycles, the original nanocrystalline region is partially converted to Zn-depleted brass and $Cu_3ZnLi_2$.

Since the electrochemical processes during Li plating and stripping operate far from thermodynamic equilibrium, there will be additional phases present, beyond the standard thermodynamically stable phases in the Cu-Zn-Li phase diagram at 300 K, including $Cu_3ZnLi_2$ [38,39]. The fully converted, homogenous grain in **Figure 4(e) –** the bottom part above the brass, noted as (g) and shown in **Figure 4(g)** is an example of such a metastable phase present within the nanocrystalline layer. Additionally, the dataset in **Figure 4(c-d)** represents likely incomplete reversion or partial decomposition of the Laves phase in the stripped nanocrystalline layer.

The nanocrystalline layer is heterogenous, as in the difference seen in **Figure 4(a)** versus **Figure 4(e)**; although admittedly speculative, the nucleated Laves phase precipitates in some areas appear to undergo classical Ostwald ripening, where the difference in chemical potential between particles of different sizes drives atoms from smaller precipitates to the larger ones. This process minimizes total interfacial energy of the system, which likely represents at least some of the Li-rich larger volumes in the upper part of **Figure *3*4(e)**'s APT dataset, as well as similar features in the upper part of **Figure 2(j)**'s APT dataset. The enlarged precipitates permanently absorb Li, as the larger precipitates do not completely electrochemically decompose. As a result, thermodynamically, the coarsening would likely lead to degraded performance over time as active Li is absorbed into and sequestered from electrochemical cycling. This creates so-called "dead Li" that becomes locked within the current collector layer, as exemplified by the middle region of the heterogeneous tip in **Figure 4(e)**, shown with the constant composition in **Figure 4(g).** The area below the reacted region is strongly Zn depleted, leading to stress build-up. Deeper lithiation accentuates the stress and drives dynamic recrystallization and the propagation of the conversion and reaction fronts downwards, as illustrated in **Figure 5(d)**.

Beyond the formation of the ternary Laves phase at grain boundaries and additional lithiation, the nanocrystalline CuZn current collector undergoes significant grain coarsening during electrochemical cycling. This coarsening is driven by multiple, interrelated factors which destabilise the initial nanocrystalline microstructure. The repeated mechanical stress from Li stripping and plating creates volumetric changes, as does Zn migration across the interlayer. These cyclic stresses, combined with compositional gradients established by both Li and Zn migration, provide both thermodynamic and kinetic driving forces for grain coarsening. The growth of larger grains, e.g. coarsening, some of which are visible in the 4D-STEM image in **Figure 3(c)**, is by itself thermodynamically favorable because they reduce the excess surface energy associated with the higher surface-to-volume ratios of nanocrystalline grains. Some grain boundary migration occurs, where larger grains consume their neighbours. Likewise, since some of the nanocrystalline grains share similar orientation as evidenced by their aligned atomic planes in **Figure 2(i)**, more clearly visible in the enlarged image in **Figure S3**, those grains can coalesce or merge. In **Figure 5(d)**, the conversion front has progressed through the original nanocrystalline zone; the stress gradients have converted the single grain underneath into nanocrystalline regions, which will continue propagating downwards, driving microstructural evolution and further dynamic recrystallization.

We acknowledge that this study only structurally investigated the initial and final (after 100 cycles) stages of microstructural evolution in the plated and stripped states. Admittedly, that data somewhat limits the ability to determine the speed of dynamic recrystallization and temporal sequence of microstructural evolution, including Ostwald ripening. However, the proposed mechanisms are consistent with the microstructural evolution seen in an analogous system under other conditions. Future work could exploit interrupted cycling at different time points, as selected by interesting electrochemical behavior, could help understand the temporal evolution of the microstructure. For instance, the electrochemical performance, as shown by the capacity retention in **Figure 1(b)**, indicates that the behaviour changed at approximately cycle 40. We interpret this as the conversion front passing the nanocrystalline layer and proceeding into the bulk. In addition, it would be important to push the understanding the role of structural and microstructural defect on the transformation as well as the kinetics of the lithiation process, along with quantifying mechanical stresses and their more specific influence on cycling performance. These aspects need further investigations that are beyond the scope of the present paper.

## Conclusion

This work reports the observation of the ternary phase, $Cu_3ZnLi_2$, in Li-metal anode-free batteries solely via electrochemistry during normal operation. The rapid Zn diffusion through the nanostructure of the electrode is accompanied by Kirkendall vacancy fluxes and resultant mechanical stresses that further modify the microstructure, and facilitate the formation of nanocrystalline $Cu_3ZnLi_2$ and its coarsening. During stripping, these ternary phases do not fully decompose, forming a Cu-rich "skeleton" and trapping Li both inside the Laves phase and in pockets related to Kirkendall voids, leading to "dead Li" and a capacity loss. These findings suggest that battery cycling with bimetallic current collectors, particularly nanostructured binary alloys, may form previously uncharacterized metastable phases during cycling that affect the Li behaviour and explain the performance evolution. Such detailed studies are needed to help guide the design of the next generation of electrodes for LMBs.

# Experimental Methods

## Materials

Sheets of alpha brass (Cu: 63% Zn 37%), hereafter CuZn37, 0.4 mm thick, were obtained from Metall Ehrnsberger (Metall Ehrnsberger GmBH, Teublitz, Germany). The material was cut into 10 mm discs and annealed for 1 hour at 600 °C under argon with furnace cooling under argon to relieve stress.

## Polishing

The discs were ground sequentially with 400, 800, 1200, 2000, and 4000 grit polishing paper at 120rpm, with a short time with a 1 µm polishing pad to create the nanocrystalline layer for investigation of mechanical surface deformation on the lithiation process.

## Sample Storage

Samples were kept in a Sylatech argon glovebox kept at less than 1ppm $O_2$ and $H_2O$ (Sylatech GB-1200E, Sylatech GmbH, Walzbachtal, Germany), equipped with a transfer port for the Ferrovac suitcase.

## X-Ray Diffraction (XRD)

A Rigaku Smartlab 9kW XRD (Rikagu Corporation, Tokyo, Japan) with a 300mm goniometer radius equipped with a Cu Kα X-ray source configured in parallel beam / grazing incidence XRD (GIXRD) mode was used to characterize the sample, set at 45 kV and 200 mA. A Goebel mirror was used with incident optics as follows: Soller slit 5.0°, 0.3mm incident slit, 5mm length limit slit. Receiving optics were as follows. 20mm receiving slit, Soller slit PSA90 0.228°, with a HyPix 3000 2D detector. Scan incidence angle was 0.75° in continuous (0D) / XRF reduction mode, with a 10° to 120° scan range with 1°/min scan speed and sampling step size 0.01°. The scans were performed in air owing to the lack of an air protection holder.

## Transmission Electron Microscopy (TEM) / Scanning Transmission Electron Microscopy (STEM)

A JEOL 2200 TEM (JEOL Ltd.) operated at 200 kV, equipped with STEM detectors, Energy Dispersive X-ray Spectroscopy (EDX), and a NanoMEGAS precession electron diffraction (PED) ASTAR system, was used for most of the TEM, EDX, and 4D-STEM analyses. For 4D-STEM of the Li-plated and Li-stripped brass specimens, the JEOL TEM was operated at 200 kV with a 10 µm condenser aperture and a 50 eV energy filter. During each 4D-STEM scan, a ~2 nm probe was rastered over a 256 nm × 256 nm area in 1 nm steps with a precession angle of 0.1°, acquiring diffraction patterns on a 4k × 4k CMOS detector (TemCam-XF416-TVIPS) with a 20 ms dwell time. The multi-index algorithm in ASTAR automatically determined the phase and orientation of overlapping grains [73,74]. The data for the stripped brass specimen was gathered using the camera in 2k x 2k mode with otherwise similar parameters. 4D-STEM indexing was performed, and only data with a reliability index of 10 or more out of 100 were included in phase plots. Scanning transmission electron microscopy (STEM)-EDS and electron energy loss spectroscopy (**Figure S4**) spectrum imaging were performed on a Titan Themis microscope operated at 300 kV, collected using a SuperX detector, a Gatan Quantum ERS energy filter, respectively. Spectrum imaging datasets were processed by multivariate statistical analysis using noise reduction [75] and background removal to highlight the Li-K edge [27].

## Focused Ion Beam (FIB) & Scanning Electron Microscopy (SEM)

A Thermo-Fisher Helios 5 FIB/SEM (Thermo Fisher Scientific, Eindhoven, Netherlands) equipped with a freely rotating Aquilos-like cryo stage and cryogenic manipulator (EZ-Lift) was used to prepare the samples, with the cryo-stage heater set to -190 °C using a $N_2$ flow rate of 190 mg/s for cryogenic specimen preparation. A lamella was extracted at cryogenic temperature and attached to the manipulator using redeposition welding, then warmed to room temperature for Pt welding to Si posts on a 36- microtip coupon (FT-36, Cameca Instruments). The cryo-stage and manipulator were cooled again for FIB sharpening at 30kV. The prepared APT specimens were cleaned with a final beam shower at 5 kV, after the protocols in [76,77]. The instrument was equipped with an EDAX Clarity direction electron detector (DED) for both Electron Backscatter Diffraction (EBSD) and Transmission Kikuchi Diffraction (TKD) (EDAX Inc., Pleasanton, CA, USA) running APEX and OIM 9.1 (beta). Transfers of Li-plated specimens under ultra-high vacuum (UHV) and/or cryogenic conditions to and from the system were done with a Ferrovac D-100 UHV cryogenic transfer suitcase (Ferrovac AG, Zurich, Switzerland), hereafter "suitcase". The bulk sample and FIB half-grid were carried between glovebox and FIB using a custom-designed Cu shuttle for the suitcase. FIB half grids were mounted in a so-called Herbig holder, which can be directly mounted in a JEOL TEM [78] to minimize loading requirements. This system is integrated with and compatible with the gloveboxes, shuttles, FIBs, reaction chamber, and APTs, as discussed in [47,48]. After initial fabrication the lamellae were stored in a desiccator. Specimens were prepared after GIXRD so were exposed to air for the time required for the scan; the XPS was performed after 30s air exposure required for loading, otherwise transferred under Ar atmosphere.

## Atom Probe Tomography (APT)

A Cameca LEAP 5000 XR reflectron atom probe (Cameca Instruments, Madison, WI) in laser mode was used for APT analysis, with the samples transferred from the FIB using the Ferrovac suitcase. Run conditions for the APT data were as follows: 50 K, 0.3 – 0.35 % detection rate (DR), 35 pJ laser pulse energy (LPE), 125 kHz acquisition rate unless otherwise specified. Data was reconstructed with a fixed shank angle; the stripped datasets used 10°. All APT images were constructed from a 15 nm thick XZ slice centered in the middle of the dataset to avoid artifacts created by voltage mode reconstruction, which should correspond to physical continuity. Scale bar is given based on reconstruction parameters and does not necessarily correspond to actual physical distance, one of the standard constraints and caveats of APT reconstructions.

### X-Ray Photoelectron Spectroscopy (XPS)

X-ray Photoelectron Spectroscopy (XPS) data was acquired on a Physical Electronics PHI Quantera II (Physical Electronics, Inc., Changhassen, MN, USA), and analysed with CasaXPS 2.3.15 (Casa Software Ltd., Devon, UK). All spectra were recorded at a 45° take-off angle. Survey scans utilized a pass energy of 112 eV and an energy step size of 0.1 eV. For detailed chemical state analysis, high-resolution scans of the O 1s, C 1s, Zn 2p and Cu 2p core levels were acquired. O1s and C1s spectra were acquired with a pass energy of 26 eV and an energy step size of 0.025 eV, whereas the Cu 2p and Zn 2p spectra were acquired with a pass energy of 55 eV and step size of 0.05 eV. To probe subsurface composition, depth profiling was performed by sputtering the sample stepwise with a 2 mm x 2 mm, 2 keV $Ar^+$ ion beam for a cumulative duration of 1158 s.

### Battery Cell Preparation

A Swagelok coin cell was configured with Li|Brass, employing 80 µL 1M $LiPF_6$, EC/DMC (1:1, v/v) electrolyte with a Celgard 2500 separator and was assembled in an Ar-filled glovebox. Three battery replicates were cycled at 0.5mA/$cm^2$, 1mAh/$cm^2$ for 100 cycles using a Neware battery cycler at room temperature, as discussed in another manuscript. After the batteries were cycled, they were disassembled in the Ar-filled glovebox, rinsed with DMC, and the cycled battery anodes were stored there. Samples were placed in the loadlock under vacuum overnight to degas before further analysis, to reduce adsorbed material. From a terminology point of view, the brass current collector was also the working electrode.

### Credit Taxonomy

E.V.W. conceived and designed the experiments, performed the majority of experimental work and data analysis, prepared the manuscript draft, and carried out project administration. J.L. and Y. Zhang contributed to battery cell fabrication, cycling, and data interpretation, and revised the manuscript. X.C. performed advanced microscopy and contributed to data analysis and manuscript preparation. J.M.P. and P.J.K. took and analyzed the XPS data and wrote that section of the manuscript; P.J.K. prepared the XPS visualizations. M.J.K. contributed to XRD data analysis, wrote that section of the manuscript, and prepared the XRD visualizations. M.P.S. and Y.Z. contributed to sample preparation and data collection. S. Zhang contributed to electron microscopy and methods. S. Zaefferer provided supervision, assisted with data analysis and validation, and revised the manuscript. Y.J. contributed to data analysis, project administration, and performed extensive manuscript editing. B.G. supervised the project, contributed to conceptualization, acquired the funding, carried out project administration, and performed extensive manuscript editing. All authors have given approval to the final version of the manuscript.

## Acknowledgements and Funding

The authors would like to thank Jürgen Wichert and the MPIE machine shop for their help in sample preparation. The authors would like to thank Christian Broß, Heidi Bögerhausen, and Katja Ängehangt for their assistance with SEM, EBSD, and sample preparation. The authors would like to thank Dr. René du Kloe from EDAX for his assistance with TKD analysis. The authors would like to thank Benjamin Breitbach for his assistance with XRD data acquisition and operation of the MPI Susmat XRD facility. The authors would like to thank Andreas Sturm for his assistance with the operation of the MPI SusMat FIB/SEM facility and Uwe Tezins for his assistance with the operation of the MPI SusMat APT facility.

E.V.W. and B.G. thank the German Research Foundation (DFG) for their generous support with the Leibniz Prize. M.J.K. thanks the Fraunhofer Society and Max Planck Society (MPS) for funding through the cooperation project MaRS ("Critical Materials Lean Magnets by Recycling and Substitution"). J.L. was supported by the Natural Sciences and Engineering Research Council of Canada (NSERC Discovery Program, RGPIN-2023-03655), Canada Foundation for Innovation (CFI), BC Knowledge Development Fund (BCKDF), and the University of British Columbia (UBC). X.C. gratefully acknowledges the DFG for support via Collaborative Research Centre/Transregio (CRC/TRR) 270 HoMMage-Z01 project. Y.J. and M.P.S. are grateful for funding from EU Horizon Research and Innovation Actions under grant agreement number 101192848 (FULLMAP). B.G. thanks the DFG for support through DIP Project No. 450800666. Y.J. and Y. Zhao acknowledge funding by the European Union through EIC grant no. 101184347 (Heat2Battery). S.Z. acknowledges DFG support through SPP 2370 (Project number 502202153). Yuwei Zhang acknowledges support from the MPS.

Funding for open access publication was provided by Project DEAL.

## Data Availability

The data is available from the authors upon reasonable request.

## Conflicts of Interest

The authors declare no conflicts of interest.

Supplementary Information

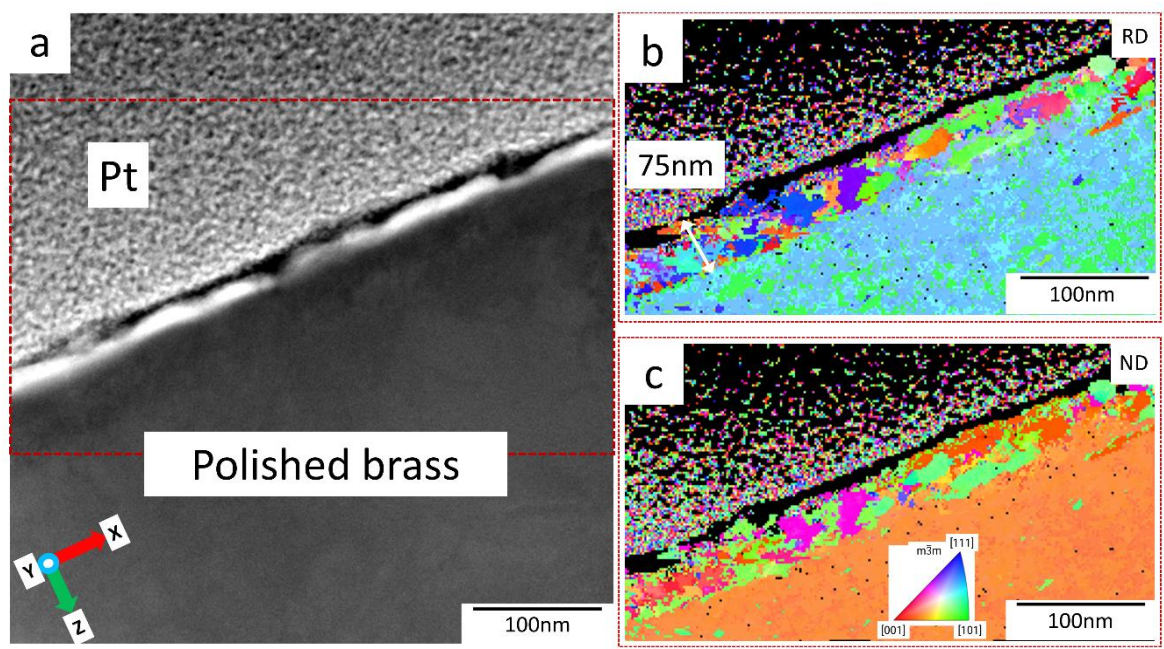

**Figure S1.** TEM of initially prepared brass foil before cycling (a) bright-field TEM view of Pt-capped polished brass lamella; note the nanocrystalline layer is not visible (b) 4D-STEM nanocrystalline grain view, rolling direction – X- direction in lamella view (c) 4D-STEM nanocrystalline grain view, normal direction, into the micrograph

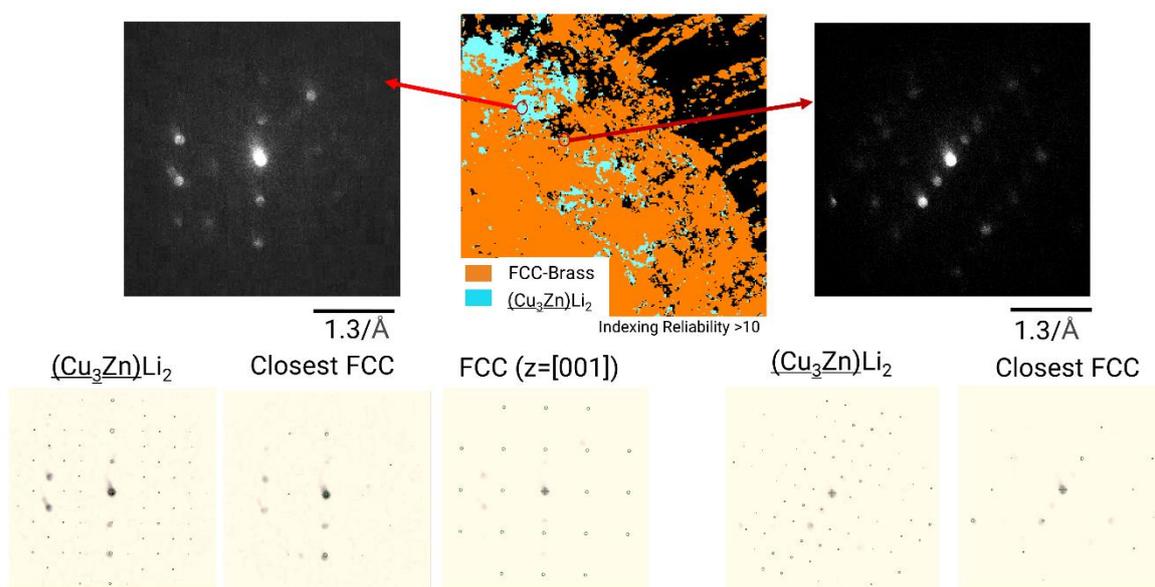

**Figure S2**. TEM indexing details, comparing Figure 2(e) and 2(f) to their closest phase matches for indexing

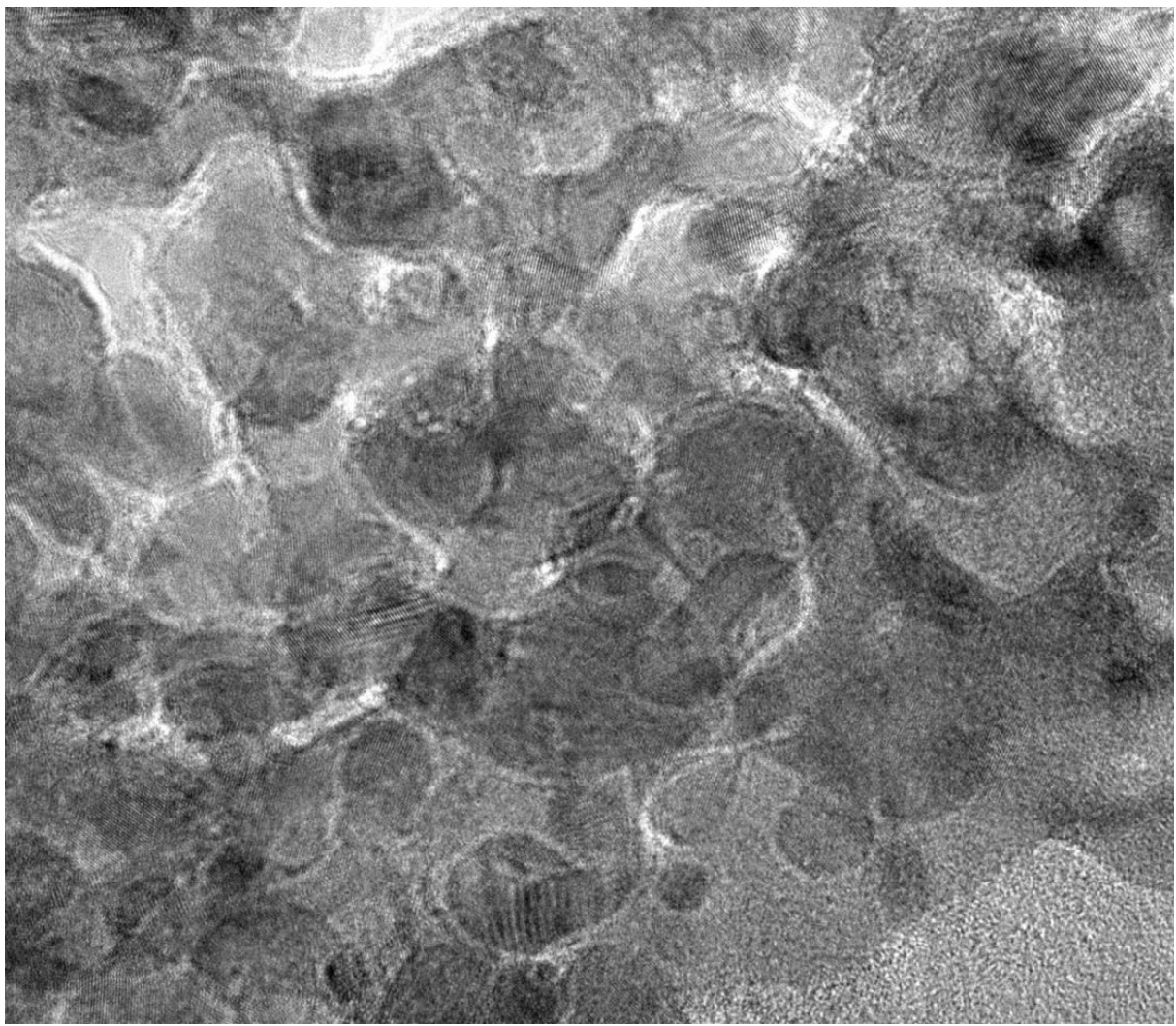

**Figure S3.** Higher-resolution view of TEM image in Figure 2, showing atomic lattice planes

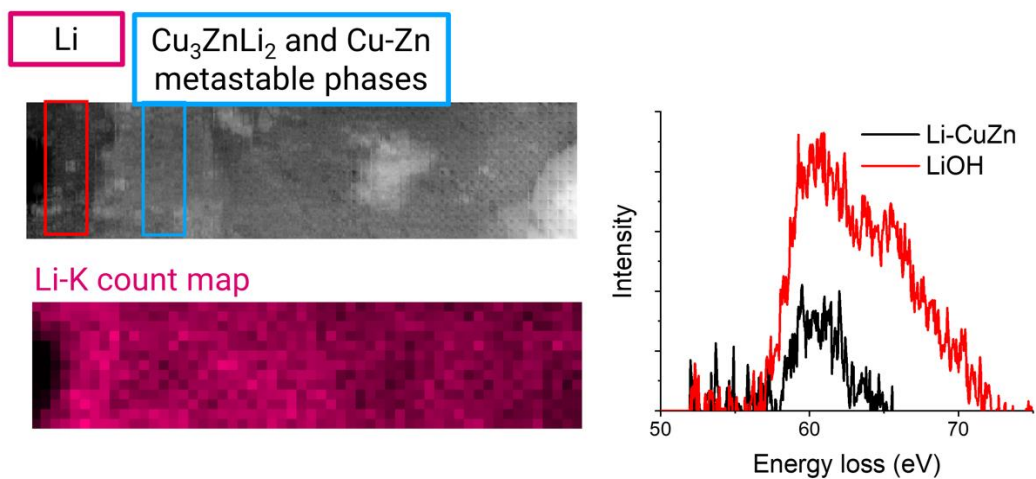

**Figure S4.** TEM lamella for plated sample; lamella was stored in deisccator so much Li was oxidized into LiOH. Top left: TEM image with labeled phases. Bottom left: Li-K count map Right: Li-oxidation state.

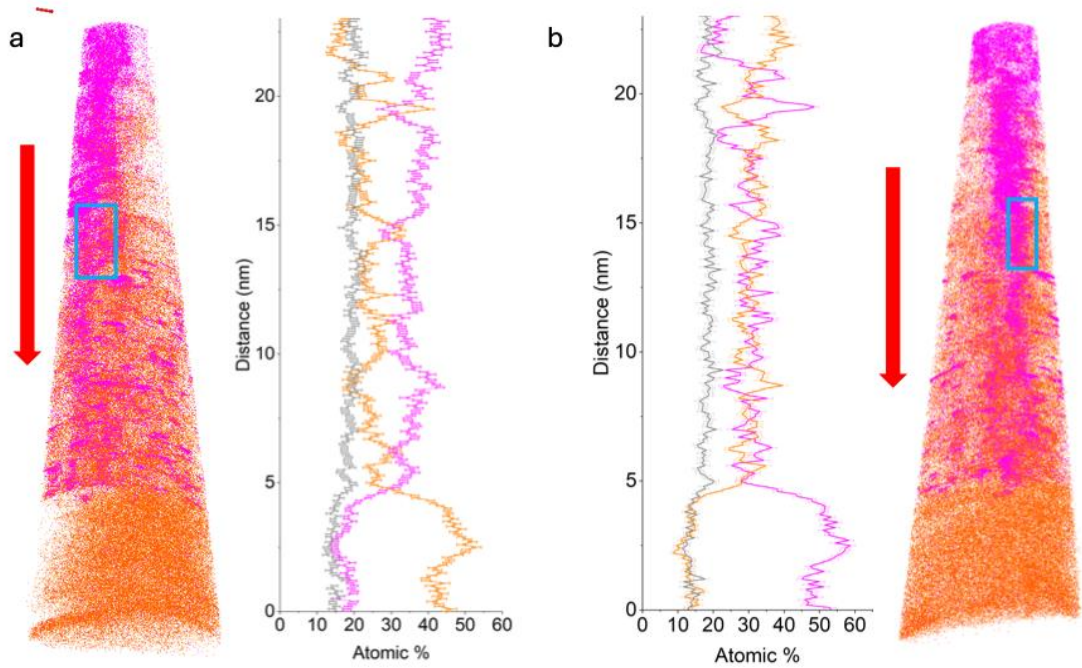

**Figure S5.** Plated Li sub-regions in area under Li-cap (a) Li-Cu 3D view with vertical 1D Z-profile showing Li-rich metastable phase in blue box (b) seond region showing Cu-Li composition equivalent 1D Z-profile in blue box ROI

**Figure S6**. XPS data analysis for electrochemically stripped 100X polished brass sample. (a) Survey scan results (b) composition of Zn, Cu, Li as a function of sputter etch time (c) Li high-resolution scan 0 s

etch time (d) Cu hi-resolution scan 0 s etch time (e) Zn high-resolution scan 0 s etch time (f) Li high-resolution scan 234 s etch time (g) Cu high-resolution scan 234 s etch time (h) Zn high-resolution scan 234 s etch time (h) Li high-resolution scan 954 s etch time (i) Cu high-resolution scan 954 s etch time (j) Zn high resolution scan 954 s etch time (k) Li high-resolution scan 2757 s etch time (l) Cu high-resolution scan 2757 s etch time (m) Zn high resolution scan 2757 s etch time (n) Li high-resolution scan 3837 s etch time (o) Cu high-resolution scan 3837 s etch time (p) Zn high resolution scan 3837 s etch time

The XPS results of sample flat polished brass (FB) Iindicate the presence of the following elements: O (*O 1s*), F (*F 1s*), C (*C 1s*), P (*P 2p*), Li (*Li 1s*), Zn (*Zn 2p$_{3/2}$*), La (*La 3d$_{5/2}$*), Cu (*Cu 2p$_{3/2}$*), Al (*Al 2p*), Ni (*Ni 2p$_{3/2}$*) and Fe (*Fe 2p$_{3/2}$*) (SM a). The depth profile of the FB sample (etching down to 3837 s at SM xx a-1 intervals) showed a decrease in the presence of *Li 1s* (from 99.52 at.% at 0 s) and an increase in *Cu 2p$_{3/2}$* and *Zn 2p$_{3/2}$*. Zn is around 18.24% (from ~0.16 at.% at 0 s), Cu 80.56% (from 0.31 at.% at 0 s) and the rest is Li (~1.2%). The detail analysis showed for in-depth profile of *Li 1s* at 0 s the presence of the Li metal binding state (~53.66 eV), Li$^{1+}$ in Li$_2$O state (~55.56 eV) and Li complex peak, which can be correlated to combination of LiF/ LiFe/LiFePO$_4$ (~57.06 eV). At 254 s of etching, the presence of Li$^{1+}$ as Li$_2$O (~55.34 eV), LiF (~55.78 eV) and Li complex state (~58.64 eV), which can be correlated to the LiFe/LiFePO$_4$. At the next selected etching step 954 s, the presence of four finding state can be observed: Li metal (~53.79 eV), Li$^{1+}$ as in Li$_2$O (~55.39 eV), LiF (~56.49 eV), Li complex (~57.98 eV), which is again contributed to the LiFe/LiFePO$_4$.https://xpsdatabase.net/. Li complex state is associated with phosphates, which are correlated to the presence of P in the sample. At the next step at 2757 s of etching the two binding states are present: Li$^{1+}$ oxidation state as in Li$_2$O (~55.62 eV) and LiF (~57.32 eV). At the last step of etching, 3837 s, the two binding state are present Li$^{1+}$ as in LiOH (~54.13 eV) suggested form and Li1+ oxidation state as in Li$_2$O (~55.71 eV) compound type. Next, the observation of *Cu 2p$_{3/2}$* in FB sample, at 0 s, shows the presence of two binding states: Cu$^{1+}$ oxidation state (~932.81 eV), which is proposed to be in Cu$_2$O binding state and Cu$^{2+}$ as in CuO (~933.71 eV) compound type. At the next etching step 234 s, three binding state are observed as Cu$^{1+}$ as suggested in Cu$_2$O (~932.02 eV), Cu$^{2+}$ as in CuO (~933.06 eV) and Cu complex state (~934.70 eV), which is suggested to be correlated to the mix presence of halides, hydroxides, and phosphates. At the next etching step at 954 s, the same three binding state are observed, and are shifted for ~0.5 eV. In the next etching interval, at 2757 s, only 2 binding states od Cu can be observed: Cu$^{2+}$ as in CuO (~933.48 eV) and Cu complex binding state (~933.78 eV), which can be a combined peak of hydroxide and halides (F). At the last etching step, 3837s, also the same peaks are present, however the complex Cu state is predominant. The observation of *Zn 2p$_{3/2}$* peak at 0 s in Fb sample, showed the presence of the following two binding state of Zn: Zn in phosphide (~1020.41 eV) and Zn$^{2+}$ as ZnO (~1021.81 eV). At the next etching step, 234 s, three binding state are present Zn phosphide (~1020.16 eV), Zn$^{2+}$ as in ZnO (~1021.86 eV) and Zn halides (~1023.16 eV). At the next etching, 954 s, also the same binding state, where slight increase of phosphide and halide state is observed. At the next etching, 2757 s, three binding states are observed, Zn in phosphide (~1020.35 eV), Zn$^{2+}$ as in ZnO (~1021.74 eV) and Zn halides state (~1022.74 eV). At the last etching step, 3837 s, Zn phosphides (~1020.59 eV), Zn$^{2+}$ as in ZnO (~1021.79 eV) and Zn halides (~1022.44 eV).

**Supplementary Table T1.** XRD phase fractions and statistics

| Phases | Plated | Stripped |
|---|---|---|
| Cu | 45.2 | 60.3 |
| Li | 16.9 | 10.8 |
| $Cu_3ZnLi_2$ | 8.1 | 6.5 |
| LiOH | 24.9 | 16.6 |
| $Zn_3P_3$ | 2 | 1.7 |
| $Zn(OH)_2$ | 0.5 | 0.5 |
| $Li_2Zn_3$ | 1.4 | 2 |
| $Cu_{0.7}Zn_2$ | 1 | 1 |

Average error is:

- For phases up to total 5 wt.% is 27% or minimum 0.3 wt.%
- For phases over 5 wt.% is 8%
- Chi-squared for plated specimen is 27.5 with amorphous or 5.5 without
- Chi-squared for stripped specimen is 24.4 with amorphous or 5.5 without

## Brief Comment on Thermodynamic Stability

The computed formation enthalpies of all candidate Li-Zn-Cu compositions at 300K, plotted against chemical potentials of Li, Zn, and Cu to construct a convex hull. The lower convex envelope identifies compositions that resist decomposition under the changing chemical environments encountered during battery cycling. Since $Cu_3ZnLi_2$ lies on the convex hull, it has negative formation enthalpy with no lower-energy decomposition pathway (see SI 12 & 13 in (*35*)).